\documentclass[aps,prd,reprint,preprintnumbers,superscriptaddress,showpacs,twocolumn]{revtex4-1}
\usepackage{latexsym,graphicx,amssymb,amsmath,mathrsfs}
\usepackage{setspace,bm}
\usepackage[colorlinks=true, pdfstartview=FitV, linkcolor=red, citecolor=blue, urlcolor=blue]{hyperref}
\usepackage[caption=false]{subfig}
\usepackage{color}

\graphicspath{{./Figures/}}

\newcommand{\diff}{\mathrm{d}}
\newcommand{\p}{\partial}

\newcommand{\Diff}{{\mathcal{D}}}

\newcommand{\be}{\begin{equation}}
\newcommand{\ee}{\end{equation}}
\newcommand{\bea}{\begin{eqnarray}}
\newcommand{\eea}{\end{eqnarray}}
\newcommand{\im}{\mathrm{i}}

\usepackage[normalem]{ulem}

\begin{document}
\preprint{RIKEN-QHP-183, RIKEN-STAMP-2, BI-TP 2015/07, YITP-15-26}

\title{Evading the sign problem in the mean-field approximation\\ through Lefschetz-thimble path integral}

\author{Yuya Tanizaki}
\email{yuya.tanizaki@riken.jp}
\affiliation{Department of Physics, The University of Tokyo, Bunkyo-ku, Tokyo 113-0033, Japan}
\affiliation{Theoretical Research Division, Nishina Center, RIKEN, Wako, Saitama, 351-0198, Japan}

\author{Hiromichi Nishimura}
\email[]{nishimura@physik.uni-bielefeld.de}
\affiliation{Faculty of Physics, Bielefeld University, D-33615 Bielefeld, Germany}

\author{Kouji Kashiwa}
\email[]{kouji.kashiwa@yukawa.kyoto-u.ac.jp}
\affiliation{Yukawa Institute for Theoretical Physics, Kyoto University,
Kyoto 606-8502, Japan}

\date{\today}

\begin{abstract}
The fermion sign problem appearing in the mean-field approximation is considered, and the systematic computational scheme of the free energy is devised by using the Lefschetz-thimble method. 
We show that the Lefschetz-thimble method respects the reflection symmetry, which makes physical quantities manifestly real at any order of approximations using complex saddle points. 
The formula is demonstrated through the Airy integral as an example, and its application to the Polyakov-loop effective model of dense QCD is discussed in detail. 
\end{abstract}


\maketitle

\section{Introduction}\label{sec:intro}
The fermion sign problem is a big obstacle for nonperturbative studies of various quantum systems ranging from condensed matter \cite{Loh:1990zz} to hadron and particle physics \cite{Muroya:2003qs}. 
The Boltzmann weight  in the path-integral expression becomes oscillatory, and importance sampling of Monte Carlo simulations fails \cite{Troyer:2004ge}.

Finite-temperature quantum chromodynamics (QCD) has been {well understood due} to the success of lattice Monte Carlo simulations
\cite{Karsch:2001cy} and of effective-model descriptions of the
confinement-deconfinement crossover \cite{Gross:1981br, Weiss:1980rj, Green:1983sd, 
Gocksch:1984yk, Meisinger:2001cq, Fukushima:2003fw}.
On the other hand, a lattice simulation of dense QCD faces the fermion sign
problem \cite{Muroya:2003qs}, which prevents us from exploring
{properties of QCD matters under extreme conditions, e.g., in colliders
and in neutron stars \cite{MeyerOrtmanns:1996ea,Wilczek:1999ym}.
Moreover, it causes a severe disease in model studies as well:
The effective action as a function of order parameters takes complex values
\cite{KorthalsAltes:1999cp, Dumitru:2005ng}. 
The evaluation of the mean-field free energy requires an oscillatory integral, and the sign problem remains \cite{Fukushima:2006uv}. 
This situation motivates us to devise a systematic and exact computational
scheme that evades the sign problem.

Recently, the Picard--Lefschetz theory turns out to be helpful for
evaluating such oscillatory integrations by using the
saddle-point analysis in the complexified space of field variables
\cite{pham1983vanishing, Witten:2010cx, Witten:2010zr, Tanizaki:2014xba,
Tanizaki:2014tua, Kanazawa:2014qma}, and it is applied to lattice Monte
Carlo simulation \cite{Cristoforetti:2012su,
Cristoforetti:2013wha, Cristoforetti:2014gsa,  Fujii:2013sra,
Mukherjee:2014hsa}. 
Other progress has been made in the Polyakov-loop effective model: Even after the analytic continuation of gauge fields, the free energy becomes real by choosing a complex saddle point that respects charge- and complex-conjugation symmetry \cite{Nishimura:2014rxa, Nishimura:2014kla}.

Unifying these ideas of recent progress, we derive a mathematical formula that allows us to systematically calculate the partition function. The formula is especially suitable for the saddle-point analysis, and we show that the formula makes the partition function
real valued at any order of approximations using complex saddle points. 
We demonstrate its usefulness through an example of the Airy integral. 
Its application to QCD is also discussed, and we show the result of the saddle-point approximation in the $SU(3)$ matrix model of heavy dense quarks.

\section{Formalism and Main Theorem}
Let us consider a multiple integration that gives the partition function,
\be
Z=\int_{\mathbb{R}^n}\diff^n x~\mathrm{e}^{-S(x)},
\label{Eq:General_Expression_Partition_Function}
\ee
{where $S(x)$ is the action functional of the real field
$x=(x_1,\ldots,x_n)$. }
By definition, the partition function
(\ref{Eq:General_Expression_Partition_Function}) for physical systems
must be a real quantity, however the Boltzmann weight $S(x)$ can be
complex in general.
One of the sufficient conditions for ensuring $Z\in\mathbb{R}$ is the
existence of a reflection symmetry $L:(x_i)\mapsto (L_{ij}x_j)$, which satisfies
$L_{ij}=L_{ji}\in\mathbb{R}$, $L^2=1$ and
\be
\overline{S(x)}=S(L\cdot x).
\label{Eq:General_CK_Symmetry}
\ee
In the case of finite-density QCD, we will see later that the charge conjugation gives this correspondence. 
The main purpose of this section is to derive the useful formula for the
mean-field analysis and its systematic improvement while keeping $Z$ real under Eq.(\ref{Eq:General_CK_Symmetry}).

For evaluating oscillatory multiple integrations, the Picard--Lefschetz
theory is powerful and its application to the path integral has recently gathered much attention \cite{pham1983vanishing, Witten:2010cx,
Witten:2010zr, Tanizaki:2014xba, Cristoforetti:2012su,
Cristoforetti:2013wha, Cristoforetti:2014gsa,  Fujii:2013sra,
Mukherjee:2014hsa, Tanizaki:2014tua, Kanazawa:2014qma}:
This method complexifies the space of field variables and finds
the most convergent integration cycle around the saddle point, which is called the Lefschetz thimble.
Let us denote complexified variables of $x$ as $z$.

To construct the Lefschetz thimble, we look for the
saddle points $z^{\sigma}$ in the complexified field space,
which are formally labeled by $\sigma\in\Sigma$. 
The directions of quantum fluctuation around $z^{\sigma}$ can be designated
by solving Morse's flow equation~\cite{pham1983vanishing, Witten:2010cx,
Witten:2010zr,Tanizaki:2014xba}:
\begin{equation}
\frac{\diff {z_i}}{\diff t} =   \overline{\left(\frac{\partial S(z)}{\partial z_i}\right)} .
\label{Eq:Downward_Flow}
\end{equation}
The Lefschetz thimble $\mathfrak{J}_{\sigma}$ is identified as the set
of points reached by some flows emanating from $z^{\sigma}$.
As a consequence, each $\mathfrak{J}_\sigma$ forms an $n$-dimensional real manifold. 
The partition function can now be computed as
\be
Z=\sum_{\sigma\in\Sigma}n_{\sigma}\int_{\mathfrak{J}_{\sigma}}\diff^n z~\mathrm{e}^{-S(z)}.
\label{Eq:Lefschetz_Thimble_Decomposition}
\ee
Therefore, the original integration
(\ref{Eq:General_Expression_Partition_Function}) is now decomposed into
the sum of the nicely converging integrations around the saddle points
$z^{\sigma}$.
The coefficient $n_{\sigma}$ is  given by the (oriented)
intersection number between $\mathbb{R}^n$ and $\mathfrak{K}_{\sigma}$;
$n_{\sigma}=\langle\mathfrak{K}_{\sigma},\mathbb{R}^n\rangle$. 
{The dual thimble} $\mathfrak{K}_{\sigma}$ is defined as the set of the points reached by
flows getting sucked into $z^\sigma$.

Let us rewrite Eq.(\ref{Eq:Lefschetz_Thimble_Decomposition}) into the
formula manifesting $Z\in\mathbb{R}$. We extend the linear map $L$ to an antilinear map
\be
K:(z_i)\mapsto (L_{ij}\overline{z_j}).
\label{Eq:Anti_Linear_Mapping}
\ee
The flow equation (\ref{Eq:Downward_Flow}) shows the covariance under
this antilinear reflection: Using Eq.(\ref{Eq:General_CK_Symmetry}), we find
\be
{\diff \overline{z_i}\over \diff t}=\overline{\left({\p S(L\cdot\overline{z})\over \p \overline{z_i}}\right)},
\ee
and thus the antilinearly transformed function $K(z(t))$ satisfies the
same flow equation (\ref{Eq:Downward_Flow}).
Many important properties can be deduced from this conclusion:
\begin{enumerate}
\item When $z^{\sigma}$ is a saddle point, so is $K({z^{\sigma}})$. Let
      $\mathfrak{J}_{\sigma}$ be the Lefschetz thimble around
      $z^{\sigma}$; then $\mathfrak{J}_{\sigma}^{K}:=\{K({z})\ | \ z\in
      \mathfrak{J}_{\sigma}\}$ coincides with the Lefschetz thimble
      around $K(z^{\sigma})$ up to orientation.
\item Since the flow is covariant under the antilinear map, $\langle
      \mathfrak{K}_{\sigma},\mathbb{R}^n\rangle = \langle
      \mathfrak{K}_{\sigma}^K,\mathbb{R}^n\rangle$ under an appropriate
      choice of the orientation.
      This means that the contributing thimbles always form an invariant pair
      $\mathfrak{J}_{\sigma}\cup\mathfrak{J}_{\sigma}^{K}$.
\end{enumerate}
{Let us decompose $\Sigma$ into three disjoint parts}. For
simplicity, we assume that $S(z^{\sigma})\in\mathbb{R}$ only if the saddle
point satisfies  $z^{\sigma}=K(z^{\sigma})$~\footnote{This
technical assumption can be removed by carefully defining $\Sigma_{\pm}$
so that they are conjugate to each other under the antilinearly extended
map $K(z)$ of $L$. }; then, $\Sigma=\Sigma_0\cup \Sigma_+ \cup \Sigma_-$,
where
\bea
\Sigma_0&=&\{\sigma\;|\; z^{\sigma}=L\cdot\overline{z^\sigma}\}, \nonumber\\
\Sigma_+&=&\{\sigma\;|\; \mathrm{Im}S(z^{\sigma})>0\},\nonumber\\
\Sigma_-&=&\{\sigma\;|\; \mathrm{Im}S(z^{\sigma})<0\}.
\eea
The transformation $K$ induces a one-to-one and onto correspondence
between $\Sigma_+$ and $\Sigma_-$. Equation
(\ref{Eq:Lefschetz_Thimble_Decomposition}) becomes
\bea
Z&=&
\sum_{\sigma\in \Sigma_0}n_{\sigma}\int_{\mathfrak{J}_{\sigma}}\diff^n z\; \mathrm{e}^{-S(z)}\nonumber\\
&+&\sum_{\tau\in\Sigma_+}n_{\tau}\int_{\mathfrak{J}_{\tau}+\mathfrak{J}_{\tau}^{{K}}}\diff^n z\; \mathrm{e}^{-S(z)}.
\label{Eq:Thimble_Decomposition_CK_Symmetry}
\eea
Each integral on the r.h.s. of the formula (\ref{Eq:Thimble_Decomposition_CK_Symmetry}) is real or purely
imaginary depending on whether $K$ changes orientation of $\mathfrak{J}_{\sigma}\cup\mathfrak{J}_{\sigma}^K$. 
Since the {l.h.s.} is real, $n_{\tau}$ must be zero unless the integral on $\mathfrak{J}_{\tau}+\mathfrak{J}_{\tau}^K$ is real
~\footnote{We here assume that the coupling constant is generic so
that Stokes phenomenon does not occur. If it does, several saddle points
are connected via some downward flows, and subdominant saddle points may
give a purely imaginary contribution so as to cancel ambiguities of the
large-order perturbation series at the dominant saddle. This is
discussed in the context of the resurgence trans-series of quantum field
theories~\cite{Unsal:2012zj, Dunne:2012ae,  Dunne:2013ada,
Cherman:2013yfa, Basar:2013eka, Cherman:2014ofa, Dorigoni:2014hea,
Larsen:2014yya}.}.
This conclusion can also be applied to expectation values of any
physical observables that satisfy the symmetry
(\ref{Eq:General_CK_Symmetry}).
The decomposition formula (\ref{Eq:Thimble_Decomposition_CK_Symmetry})
takes a suitable form for the saddle-point analysis. 

Let us emphasize here that the reality of the partition function is
ensured only by the invariance under the antilinear map $K$.
The point of our discussion is that the Lefschetz-thimble decomposition of the integration cycle manifestly respects the antilinear reflection $K$, and so does the saddle-point analysis based on it. 

There are two possible remnants of the sign problem in the Lefschetz-thimble method \cite{Aarts:2013fpa,
Aarts:2014nxa, Cristoforetti:2014gsa,  Fujii:2013sra}.
One is the complex phases coming from the Jacobian of $\diff^n z$, which is called the residual sign problem.
In Eq.(\ref{Eq:Thimble_Decomposition_CK_Symmetry}), the imaginary part of the Jacobian automatically cancels under the antilinear symmetry $K$, but its real part can flip the sign and gives the sign problem when one goes beyond the saddle-point approximation. 
Another one comes from the overall signature of each integral in Eq.(\ref{Eq:Thimble_Decomposition_CK_Symmetry}), or the signature of $n_{\sigma}$. 
A cancellation among different saddle-point contributions is important for the study of the phase transition \cite{Kanazawa:2014qma}, because the partition function vanishes at the phase transition \cite{PhysRev.87.410}. 
This can give information about the phase transition even within the saddle-point analysis.

\section{Example: Airy integral}
Let us {work on} the simplest example in order to understand the formula
(\ref{Eq:Thimble_Decomposition_CK_Symmetry}).
We consider the Airy integral with a real parameter $a$:
\be
Z_{\mathrm{Airy}}(a)={1\over 2\pi}\int_{\mathbb{R}}\diff x~\mathrm{e}^{\im\left({x^3\over 3}+a x\right)}.
\label{Eq:Airy_Integral}
\ee
Denoting the exponent of the integrand as $-S(x,a)$,
$\overline{S(x,a)}=S(-x,a)$ when $x$ and $a$ are real, and
(\ref{Eq:General_CK_Symmetry}) is satisfied. The antilinear map $K$ is
then $K(z)=-\overline{z}$.

If $a>0$, then the saddle points are $z^\sigma=\pm \sqrt{a} \im$, and
both of them are invariant under $K$; $z^{\sigma}=-\overline{z^\sigma}$.
Studying the flow equation, only one saddle point $+\sqrt{a}\im$ turns
out to contribute to Eq.(\ref{Eq:Airy_Integral}).
Using the Lefschetz-thimble decomposition
(\ref{Eq:Thimble_Decomposition_CK_Symmetry}), the saddle-point
approximation gives
\be
Z_{\mathrm{Airy}}(a)\simeq {a^{-1/4}\over 2\sqrt{\pi} } \exp
\left(-{2\over 3}a^{3/2}\right),
\label{Eq:Airy_Asymptotic_Positive}
\ee
which is the asymptotic formula of the Airy function as $a\to+\infty$
(see Chap.19 of \cite{NIST:DLMF}).
On the other hand, when $a<0$, the saddle points form the conjugate pair
under $K$, $\pm \sqrt{|a|}$, and both of them  give contribution to the
original integration.
Similarly, we get 
\be
Z_{\mathrm{Airy}}(a)\simeq {|a|^{-1/4}\over \sqrt{\pi} }\cos \left({2|a|^{3/2}\over 3}-{\pi\over 4}\right),
\label{Eq:Airy_Asymptotic_Negative}
\ee
which is nothing but the asymptotic formula of the Airy function as
$a\to-\infty$ (see Chap.19 of \cite{NIST:DLMF}). Notice that both approximate
results [Eqs. (\ref{Eq:Airy_Asymptotic_Positive}) and
(\ref{Eq:Airy_Asymptotic_Negative})] are manifestly real, as we have
already shown in general.

Let us remark that only by using $\overline{S(x,a)}=S(-x,a)$ is the
partition function ensured to be real:
\be
	Z_{\mathrm{Airy}}={1\over \pi}\int_{0}^{\infty}\diff x \cos \left({x^3\over 3}+a x\right). 
\label{Eq:Def2_Airy_Function}
\ee
However, this integration is oscillatory and still has the sign problem. 
The Lefschetz-thimble decomposition (\ref{Eq:Thimble_Decomposition_CK_Symmetry}) plays a pivotal role in evaluating these asymptotic formulas (\ref{Eq:Airy_Asymptotic_Positive}) and (\ref{Eq:Airy_Asymptotic_Negative}) without encountering the sign problem.

\section{Application to the sign\\ problem of Dense QCD}
The QCD partition function at temperature $T=\beta^{-1}$ and
quark chemical potential $\mu_{\mathrm{qk}}$ is given by
\be
Z_{\mathrm{QCD}}=\int\Diff A\; \mathrm{det}\,\mathcal{M}(\mu_{\mathrm{qk}},A)\;\mathrm{e}^{-S_{\mathrm{YM}}[A]},
\ee
where $S_{YM}={1\over 2}\mathrm{tr}\int_0^{\beta}\diff x^4\int\diff^3 \bm{x} |F_{\mu\nu}|^2\;(>0)$ is the Yang-Mills action, and
\be
\mathrm{det}\mathcal{M}(\mu_{\mathrm{qk}},A)=\mathrm{det}\left[\gamma^{\nu}(\p_{\nu}+\im g A_\nu)+\gamma^4 \mu_{\mathrm{qk}} +m_{\mathrm{qk}}\right]
\ee
is the quark determinant.
When $\mu_{\mathrm{qk}} \neq 0$, the quark determinant becomes an oscillatory functional of the gauge field $A$, and the sign problem emerges. 
Even when $\mu_{\mathrm{qk}}\not=0$, the charge conjugation~$A\mapsto-A^t$ with the $\gamma_5$ hermiticity implies that the fermion determinant still satisfies the identity~\cite{Dumitru:2005ng,Fukushima:2006uv},
\bea
\overline{\det\mathcal{M}(\mu_{\mathrm{qk}},A)}&=&\det\mathcal{M}(-\mu_{\mathrm{qk}},A^{\dagger}) \nonumber\\
&=&\det\mathcal{M}(\mu_{\mathrm{qk}},-\overline{A}). 
\label{Eq:Fermion_CK_Symmetry}
\eea
The charge $\mathcal{C}$ and complex $\mathcal{K}$ conjugation, or the $\mathcal{CK}$ transformation, serves as the antilinear map (\ref{Eq:Anti_Linear_Mapping}) for finite-density QCD \cite{Nishimura:2014rxa, Nishimura:2014kla}. 
The Lefschetz-thimble decomposition (\ref{Eq:Thimble_Decomposition_CK_Symmetry})
manifestly respects the $\mathcal{CK}$ symmetry so that $Z_{\mathrm{QCD}}\in\mathbb{R}$.
Since the discussion is robust, this conclusion applies to any effective descriptions of QCD, including lattice QCD simulations \cite{Cristoforetti:2012su, Cristoforetti:2013wha, Cristoforetti:2014gsa,  Fujii:2013sra, Mukherjee:2014hsa} and also matrix models.

We apply our insight on the Lefschetz-thimble method to the sign problem in Polyakov-loop effective models.
The Polyakov loop $\ell_{\bm{3}}$ of the fundamental representation $\bm{3}$ is an order parameter for the confinement-deconfinement transition \cite{Polyakov:1978vu};
\be
\ell_{\bm{3}}={1\over 3}\mathrm{tr}\left[\mathcal{P}\exp\left( \im g \int_0^{\beta}A_4\diff x^4\right) \right], 
\ee
where $\mathcal{P}$ refers to the path ordering. 
To understand its properties, the constrained effective action $S_{\mathrm{eff}}(a_4)$ of the Polyakov loop is useful \cite{KorthalsAltes:1993ca, Fukuda:1974ey}. 
It describes the (complex) weight of the partition function under the background Polyakov-loop phase $a_4$:
\be
\exp\left(-S_{\mathrm{eff}}(a_4)\right)=\int \Diff A \;\mathrm{e}^{-S[A]} \ \delta\left({A_4}-a_4\right),
\ee
where ${A_4}$ is the temporal gauge field and $S[A]=S_{\mathrm{YM}}[A]-\ln\mathrm{det}\mathcal{M}(\mu_{\mathrm{qk}},A)$. We
implicitly take the Polyakov gauge fixing with
\be
\exp\Bigl(\im\frac{g a_4}{T}\Bigr) = \mathrm{diag}\left[\mathrm{e}^{\im(\theta_1+\theta_2)},\mathrm{e}^{\im(-\theta_1+\theta_2)},\mathrm{e}^{-2\im\theta_2}\right].
\label{Eq:Definition_Polyakov_Gauge}
\ee
The Weyl group action $(\theta_1,\theta_2)\mapsto(-\theta_1,\theta_2)$, $(\theta_1,\theta_2)\mapsto ((\theta_1+3\theta_2)/2,(\theta_1-\theta_2)/2)$ just permutes eigenvalues of  the Polyakov loop (\ref{Eq:Definition_Polyakov_Gauge}), and
thus the parameter region can be restricted to
$\mathfrak{C}=\{\theta=(\theta_1,\theta_2)\;|\; 3|\theta_2|\le
\theta_1\le \pi\}$. 
This background field method, or the mean-field approximation (MFA), gives a single integration over $a_4(\theta_1,\theta_2)$ to compute the partition function: 
\begin{equation}
Z_{\mathrm{QCD}} = \int_{\mathfrak{C}} \diff \theta_1 \diff \theta_2  H(\theta_1,\theta_2)\exp \left[ - S_{\mathrm{eff}} (\theta_1, \theta_2) \right],
\label{Eq:Definition_Partition_Function}
\end{equation}
with $H(\theta)=\sin^2\theta_1\sin^2({(\theta_1+3\theta_2)/2}) \sin^2({(\theta_1-3\theta_2)/ 2})$ coming from the $SU(3)$ Haar
measure \cite{Green:1983sd,Meisinger:2001cq}.  
When the quark chemical potential $\mu_{\mathrm{qk}}$ is turned on under
the nontrivial Polyakov-loop background, the effective action 
$S_{\mathrm{eff}}(\theta)$ takes complex values in general due to the
quark determinant.
This makes the integration (\ref{Eq:Definition_Partition_Function}) oscillatory, and the sign problem remains in the MFA \cite{Dumitru:2005ng, Fukushima:2006uv}. 

In order to evade the sign problem of the MFA, we should rewrite Eq.(\ref{Eq:Definition_Partition_Function}) based on the decomposition formula (\ref{Eq:Thimble_Decomposition_CK_Symmetry}). 
Since $H(\theta)$ identically vanishes at the edge of $\mathfrak{C}$, the Lefschetz-thimble method can be applied by regarding
$S(\theta)=S_{\mathrm{eff}}(\theta)-\ln H(\theta)$. 
In terms of the Polyakov-loop (\ref{Eq:Definition_Polyakov_Gauge}), the
$\mathcal{CK}$ transformation $K$ becomes
\be
K(z_1,z_2)=(\overline{z_1},-\overline{z_2}),
\ee
where $z_i$ is the complexified variable of $\theta_i$, up to some Weyl
transformations.
The $\mathcal{CK}$ symmetry (\ref{Eq:Fermion_CK_Symmetry}) leads
\be
\overline{S_{\mathrm{eff}}(z_1,z_2)}=S_{\mathrm{eff}}(\overline{z_1},-\overline{z_2}).
\label{Eq:CK_Symmetry}
\ee
It is important to remark that the Polyakov loop,
\be
\ell_{\bm{3}}(\theta)={1\over 3}\left(2\mathrm{e}^{\im \theta_2}\cos\theta_1 + \mathrm{e}^{-2\im\theta_2}\right),
\ee
satisfies the $\mathcal{CK}$ symmetry (\ref{Eq:CK_Symmetry}).
For our demonstration, we take a simplified heavy-quark model~\cite{Dumitru:2005ng,
Alexandrou:1998wv, Condella:1999bk, Alford:2001ug, Banks:1983me,
Pisarski:2000eq, Dumitru:2000in}:
\bea
S_{\mathrm{eff}}&=&-h{(3^2-1)\over 2}{\Big(}e^{\mu}\ell_{\bm{3}}(\theta_1,\theta_2) +e^{-\mu}\ell_{\overline{\bm{3}}}(\theta_1,\theta_2) \Big{)}\nonumber\\
&=&-{8h\over 3}{\Big(} 2\cos\theta_1\cos(\theta_2-\im\mu)+\cos(2\theta_2+\im \mu){\Big)}.\quad {}
\eea
When $h\not=0$ and $\mu=\beta\mu_{\mathrm{qk}}\not=0$, the integration
(\ref{Eq:Definition_Partition_Function}) is oscillatory because
$S_{\mathrm{eff}}$ takes complex values.
The gluon dynamics is neglected just for simplicity. 

If $h=0$, the system has the $\mathbb{Z}_3$ symmetry generated by
$(\theta_1,\theta_2)\mapsto (\theta_1,\theta_2+2\pi/3)$ up to Weyl group
actions. The Haar measure factor $H(\theta_1,\theta_2)$ takes the unique
maxima at $\theta^*=(2\pi/3,0)$ in $\mathfrak{C}$, and this is
$\mathbb{Z}_3$ invariant.
For later purposes, it is important to notice that this saddle point is
$\mathcal{CK}$ symmetric in the sense that $\theta^*=K(\theta^*)$.
The eigenvalues of the temporal Wilson line
(\ref{Eq:Definition_Polyakov_Gauge}) become uniformly separated, and
thus $\langle \ell_{\bm{3}}\rangle=0$.

\begin{figure}[t]
\centering
\includegraphics[scale=0.4]{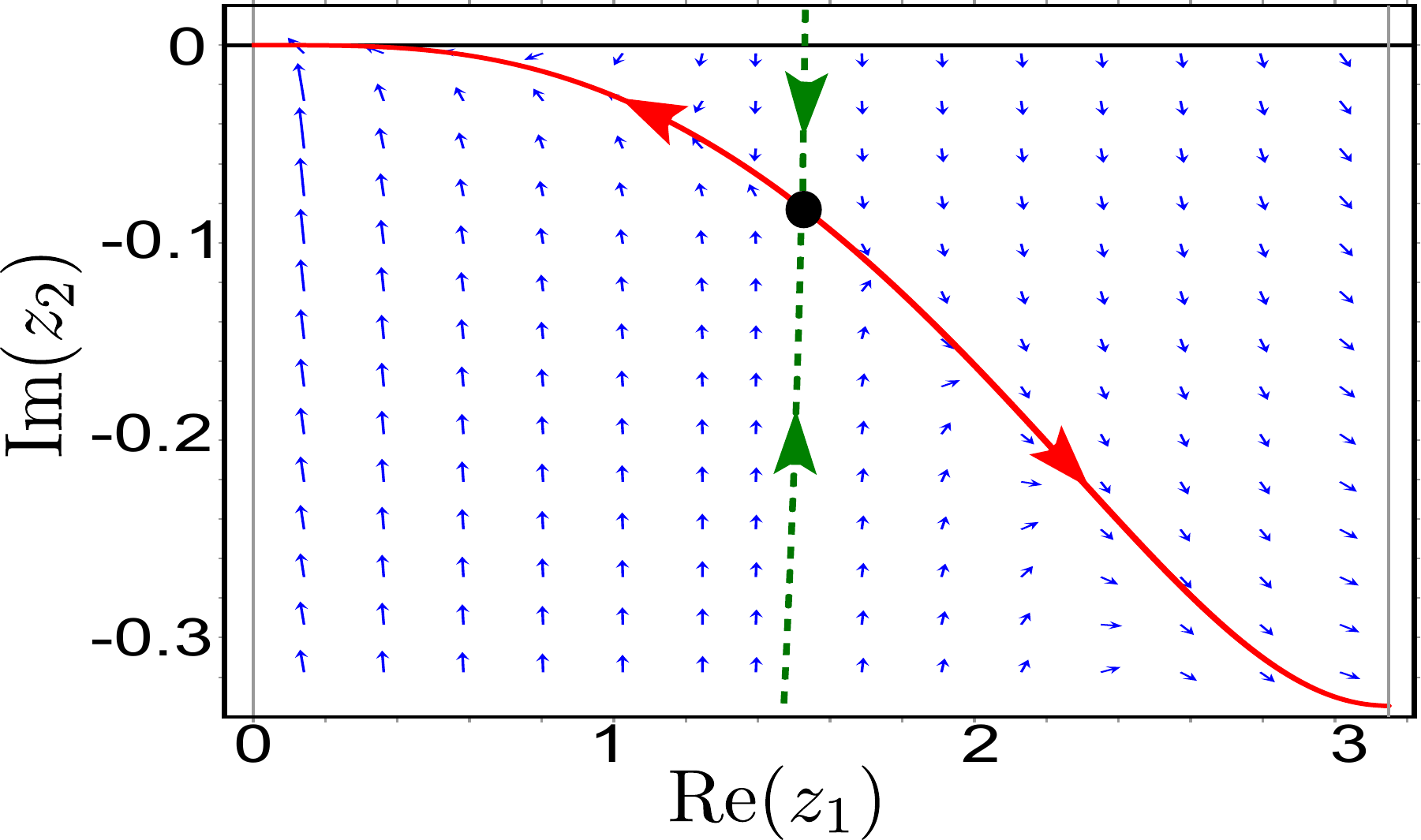}
\caption{Flows at $h=0.1$ and $\mu=2$ around the $\mathcal{CK}$-symmetric saddle point (the black blob) $z^*$ in the $\mathrm{Re}(z_1)$-$\mathrm{Im}(z_2)$ plane.  The red solid and green dashed lines are the Lefschetz thimble $\mathfrak{J}_*$ and its dual $\mathfrak{K}_*$ of $z^*$, respectively. Since $\mathfrak{K}_*$ intersects with $\mathrm{Im}(z_2)=0$, the integral over $\mathfrak{J}_*$ is equal to that over $\mathfrak{C}$. }
\label{Fig:Flow_CK_Symmetric_Plane}
\end{figure}

When $h\not=0$, the $\mathbb{Z}_3$ symmetry is explicitly broken, and the saddle point $z^*$ moves away from the $\mathbb{Z}_3$-symmetric point $(2\pi/3,0)$, but it still must be $\mathcal{CK}$ symmetric because there are no other saddles to form a pair. 
By continuity of intersection numbers, $z^*$ will contribute to $Z_{\mathrm{QCD}}$ even for finite $h$ and $\mu$. 
In order to demonstrate it lucidly, we explicitly solve the flow equation (\ref{Eq:Downward_Flow}) around the $\mathcal{CK}$-symmetric saddle point (see Fig.~\ref{Fig:Flow_CK_Symmetric_Plane}). 
The dual thimble $\mathfrak{K}_*$ of  $z^*$ is shown with the green dashed curve, and it indeed intersects with the original integration cycle $\mathfrak{C}$. 
Therefore, the complex saddle point contributes, but physical quantities take real values since the $\mathcal{CK}$ symmetry is respected under the Lefschetz-thimble decomposition. 

In the limit $\mu\to +\infty$, the saddle-point approximation becomes accurate. 
In this limit, the effect of the quark potential $S_{\mathrm{eff}}$ becomes quite large. 
By solving the saddle-point equation of $S$ in the limit $\mu\to \infty$, we find that 
\bea
z_1^* \simeq {3e^{-\mu/2}\over 2\sqrt{h}}, \quad z_2^* \simeq -\im{e^{-\mu}\over 8 h}, 
\eea
and it indeed approaches the perturbative vacuum $z^*_{\mathrm{pert.}}=(0,0)$ of $S_{\mathrm{eff}}$. 
Thus, $\langle \ell_{\bm{3}}\rangle$ and $\langle\ell_{\overline{\bm{3}}}\rangle$ converge to $1$ in this limit, as was numerically observed in Fig.2 of Ref.~\cite{Dumitru:2005ng}. 
Using the saddle-point approximation, we can see that these Polyakov loops have different expectation values: 
\bea
\langle \ell_{\overline{\bm{3}}}\rangle-\langle \ell_{\bm{3}}\rangle
&\simeq& {2\over 3}\left(\sinh 2\im z_2^*-2\cos z_1^* \sinh \im z_2^*\right)
> 0. 
\eea
These results are consistent with those of exact computations of expectation values \cite{Dumitru:2005ng}, and the $\mathcal{CK}$-symmetric saddle point gives real expectation values \cite{Nishimura:2014rxa,Nishimura:2014kla}. 
The formula (\ref{Eq:Thimble_Decomposition_CK_Symmetry}) shows that this approximation can be systematically improved perturbatively to satisfy these physical requirements. 

\section{Conclusion and Discussion}
In this paper, we have shown that the Lefschetz-thimble decomposition manifestly respects the ${\cal C K}$ symmetry of dense QCD, so that $Z_{\mathrm{QCD}}$ becomes real even with approximations. 
The Lefschetz-thimble decomposition is suitable for the saddle-point analysis, so the partition function can be computed in a systematic way. 
This solves the sign problem appearing in the MFA when the classical action is complex.
The importance of such antilinear reflection symmetry has also been discussed in the literature of $\mathcal{PT}$-symmetric quantum theory and the reality of the partition function was also ensured there \cite{Bender:1998ke,Bender:2007nj, Meisinger:2012va}. It will be interesting to establish a clear and explicit connection between these two consistent results. 

Our method was demonstrated by applying it to the Airy integral and to the $SU(3)$ matrix model of heavy dense quarks at finite $\mu$. 
Since the derivation is robust, our result can be broadly applied: In condensed matter physics, studying  Hubbard models with Hubbard-Stratonovich transformation must be an interesting example \cite{Batrouni:1992fj}, and in hadron physics 
other effective models of QCD---such as Polyakov-loop extended Nambu--Jona-Lasinio model \cite{Fukushima:2003fw}---should also be studied based on the formula (\ref{Eq:Thimble_Decomposition_CK_Symmetry}). 
There has been empirical proposals to circumvent the sign problem in the MFA \cite{Dumitru:2005ng, Fukushima:2006uv, Ratti:2005jh, Roessner:2006xn}, 
and the reality of the free energy at the $\mathcal{CK}$-symmetric saddle point has been argued in Refs.~\cite{Nishimura:2014rxa,Nishimura:2014kla}. 
The current work goes beyond the saddle-point approximation and gives an exact expression for the partition function using the Lefschetz-thimble technique (\ref{Eq:Thimble_Decomposition_CK_Symmetry}). 
This approach provides a systematic way to find relevant saddle points, and we now have a solid mathematical foundation for those approximations. 
This work will give clues for investigating the sign problem of lattice field theories. 

We should note that, however, the positivity of the partition function is not yet ensured in general cases, especially when many saddle points contribute to the partition function. 
This possible cancellation among contributions of several saddle points will be physically important in the context of phase transitions. 
In order to deepen our understanding of positivity and also of unitarity, we need to combine our result of this paper with other properties of physical systems, and we expect that it might be related to the reflection positivity \cite{osterwalder1973axioms}. 
This would be an important direction for future studies aimed at solving the sign problem of other complicated systems like chiral random matrix models \cite{Stephanov:1996ki}.

\begin{acknowledgments}
The authors thank Yoshimasa~Hidaka for reading an early stage of the manuscript and giving useful comments. 
Y.~T. acknowledges helpful discussions with Yoshimasa~Hidaka and
 Robert~D.~Pisarski.
H.~N. thanks Mike Ogilvie for useful discussions.
K.~K. thanks Hiroaki~Kouno and Masanobu~Yahiro for
 fruitful discussions.
{Y.~T. and K.~K. are supported by Grants-in-Aid for Japan Society for the Promotion
 of Science (JSPS) fellows No.25-6615 and No.26-1717, respectively. }
The work of Y.~T. was also supported by the RIKEN interdisciplinary Theoretical Science (iTHES) project, by the JSPS Strategic Young Researcher Overseas Visits Program for Accelerating Brain Circulation, and by the Program for Leading Graduate Schools, Ministry of Education, Culture, Sports, Science and Technology (MEXT) in Japan. {H.~N. is supported by Bielefeld Young Researcher's Fund.}

\end{acknowledgments}

\appendix

\bibliography{lefschetz,./ref}

\end{document}